\documentstyle[12pt]{article}

\begin{document}
\title{Remarks on the $W$ propagator at the resonance}
\author{{\bf G. L\'opez Castro$^1$, J. L. Lucio M.$^2$ and J. Pestieau$^3$} \\
$^1$ \small Departamento de F\'\i sica, Cinvestav del IPN \\
 \small Apartado Postal 14-740, 07000 M\'exico, DF. MEXICO \\
 $^2$ \small Instituto de F\'\i sica, Universidad de Guanajuato \\
 \small Apartado Postal E-143, 37150 Le\'on, Guanajuato. MEXICO \\
 $^3$ \small Institut de Physique Th\'eorique, Universit\'e \\
 \small Catholique de Louvain, Louvain-la-Neuve. BELGIUM}
\setcounter{page}{0}
\maketitle
\begin{abstract}
  We address the problem of properly defining the $W^{\pm}, \ Z^0$ propagator
   in the resonance region. Particular attention is paid to the
   longitudinal piece of this propagator. We also discuss the related
   renormalization procedures and the unitarity property.
\end{abstract}
\


\newpage
\vspace{4cm}

\

   There has been recently a renewed interest regarding the correct form
   of the massive gauge-boson propagators to be used in the resonance region
    [1-9]. In particular, the accuracy attained in LEP experiments regarding
    the $Z^0$ mass, has raised the question whether the extracted value of the
    renormalized on-shell $Z^0$ mass [10] is gauge-dependent in higher orders
of
    perturbation theory [1,4,5]. Thus, it was proposed [1,2,4,5] to return
    to the definition of the mass and width of the resonance in terms of
    the real and imaginary parts of the pole position of the amplitude. On
    this side, it has been shown [1,4,5] that a Laurent expansion around
    the complex pole, provides a systematic way to maintain the
gauge-invariance
    of the amplitude at any order of perturbation theory.

      The situation concerning the $W^{\pm}, \ Z^0$ bosons has been also
      discussed in references [3,7-9]. In some sense, the $W^{\pm}$ case is
      less complicated as far as effects related to the $\gamma Z$ mixing are
      not present; however, the $q_{\mu}q_{\nu}$ piece in the $W^{\pm}$
      propagator can
      have important effects, in contrast to the $Z^0$ case which
      usually appears coupled to
      light fermions. For instance, since $m_t > M_W +m_b$ one can easily
realize
      from references [8,9,11,12] that the correct form of the $W^{\pm}$
      propagator, and in particular of the $q_{\mu}q_{\nu}$ piece, is important
      in order to assess the size for the CP asymmetry arising from
      interference effects between two {\em top} quark decay diagrams, one
      containing a resonant $W$ propagator and the other involving a CP
      violating phase.

        In a recent paper [3] we have proposed that the correct form of the
        lowest order $W^{\pm} (Z^0)$ renormalized propagator, in the unitary
        gauge, to be used in the resonance region should be
        \vspace{.5cm}
        \begin{equation}
        \overline{\Delta}_{\mu\nu}(q)=\frac{i\left \{
        \textstyle -g_{\mu\nu} +
        q_{\mu}q_{\nu}/(M_W^2-iM_W\Gamma_W)\right \} }{
        \textstyle q^2-M_W^2+iM_W\Gamma_W},
\vspace{.5cm}
\end{equation}
       where the mass $M_W$ and width $\Gamma_W$ of thee $W$ boson are related
       (see Eq.(8) below) to the real and imaginary parts
       of the pole position, which is a basic property of $S$-matrix [13]. Our
       argument was based [3] in the fulfillment of the (lowest order) Ward
identity
       involving the electromagnetic vertex $WW\gamma$.
       As it was mentionned in ref.[3], this identity assures that the
       amplitude for processes such as $e^+e^- \rightarrow W^+W^- \rightarrow
       4{\rm fermions} +\gamma$ or $t \rightarrow b l^+ \nu_l \gamma$ are
gauge-
       invariant under electromagnetism only by using Eq.(1) above (see
also Appendix B). Furthermore, we
       mentionned [3] that to deal with an arbitrary $\xi$-gauge, the
replacement
       $M_W^2 \rightarrow M_W^2 - iM_W\Gamma_W$ should be done everywhere the
$W$
       mass appears in the usual Feynman rules [14].

       Using a different line of arguments, the authors in refs. [7] get a
       propagator similar to Eq.(1). However, in refs. [8,9] a different
conclusion is obtained. As far as the derivation in
       [3] is concerned, the authors in ref. [9] pointed out that our argument
is
       not consistent, because the electromagnetic Ward identity used in our
paper
        seems to involve the propagators and the $WW\gamma$ vertex at different
        orders.

          By using a general $\xi$-gauge we show in this paper that, when
          taken at lowest order, the
          renormalized $W^{\pm}$ propagator obtained from the Dyson summation
          indeed reproduces the resonant form suggested in refs. [3,7,8]
          . This result is obtained from the
          renormalized propagator through its Laurent expansion around the pole
          position; the non-resonant terms arising in this expansion are
          shown to be explicitly
          of higher orders in the relevant coupling constant. We also address
some
          comments on renormalization and unitarity.

          Let us start by setting our conventions. In the general $\xi$-gauge,
          $\xi$=1,0 and $\infty$ correspond to the Landau, Feynman-'t Hooft and
          unitary gauges, respectively. The bare $W$ boson propagator is given
by
          [14]:
          \vspace{.5cm}
          \begin{equation}
          P_{\mu\nu}^{(0)}(q)= \frac{i\left \{ \textstyle
-g_{\mu\nu}+(1-\xi)q_{\mu}
          q_{\nu}/(q^2-\xi M^2)\right \} }{q^2-M^2},
\vspace{.5cm}
\end{equation}
          where $M$ denotes the bare $W$ mass.

            The unrenormalized $W$ self-energy can be decomposed into different
            forms as follows:
            \vspace{.5cm}
            \begin{eqnarray*}
            i\Pi_{\mu\nu}(q) &=& \left (g_{\mu\nu}-\frac{q_{\mu}q_{\nu}}{q^2}
            \right )\Pi^{WW}_T(q^2)+\frac{q_{\mu}q_{\nu}}{q^2}\Pi^{WW}_L(q^2)
              \hspace{3.5cm} (3a) \\
            &=& -g_{\mu\nu}\Pi_T(q^2)+q_{\mu}q_{\nu}\Pi_L(q^2)
            \hspace{6.2cm} (3b) \\
            &=& g_{\mu\nu}F_1(q^2)+(q^2g_{\mu\nu}-q_{\mu}q_{\nu})F_2(q^2)
            \hspace{5cm} (3c)\\
            &=& \left ( -g_{\mu\nu}+{\frac{\textstyle
q_{\mu}q_{\nu}}{\textstyle q^2}}
            \right )\overline{\epsilon}_T(q^2)+\frac{\textstyle
q_{\mu}q_{\nu}}{\textstyle
            q^2}\overline{\epsilon}_L(q^2) \hspace{4.5cm} (3d)
\vspace{.5cm}
             \end{eqnarray*}

             where the Eqs.(3a)-(3d) are the parametrizations used
             in refs.[15,7,8,9], respectively. Note that in ref.[9] only the
             imaginary parts
             of $\overline{\epsilon}_{T,L}$ have been considered. For
simplicity, (unless specified)
              in the following we will not write the $q^2$-dependence in the
scalar
              self-energy functions.

             By choosing the first parametrization ---Eq. (3a)---, the infinite
sum of
             the 1PI bubble graphs give rise to the following full
unrenormalized
             propagator [15]:
             \setcounter{equation}{3}
             \vspace{.5cm}
             \begin{eqnarray}
             P_{\mu\nu}(q)=\frac{i}{q^2-M^2-\Pi_T^{WW}}&&\left \{-g_{\mu\nu} +
\frac{
             \textstyle q_{\mu}q_{\nu}}{\textstyle q^2} \cdot \frac{\textstyle
q^2 -
             \xi (q^2+\Pi_L^{WW}-\Pi_T^{WW})}{\textstyle q^2-\xi M^2 -\xi
\Pi_L^{WW}}
             \right. \nonumber \\ & & \left. \cdot \left [ \frac{\textstyle
1-\frac{\textstyle a}
             {\textstyle (1-\xi )q^2+\xi (\Pi_T^{WW}-\Pi_L^{WW})}}{\textstyle
1-\frac{
             \textstyle a}{\textstyle q^2-\xi M^2-\xi \Pi_L^{WW}}} \right ]
\right \}
\vspace{.5cm}
\end{eqnarray}
             where $a\equiv \left ( \frac{\textstyle \xi q^2}{\textstyle M^2}
             \right )\cdot (\Pi^{W\phi})^2
             /(q^2-\xi M^2 - \Pi^{\phi\phi})$, and $\Pi^{W\phi}, \
\Pi^{\phi\phi}$ are
             the corresponding self-energies for the $W-\phi$ and $\phi-\phi$
             fields [15]. $\phi$ is the would-be goldstone associated to the
             $W$ boson.

               It can be easily shown that the contributions of the would-be
goldstones
               (the term within squared brackets in Eq.(4)) can be written as
follows:
  \vspace{.5cm}
   \begin{equation}
               1-\frac{\textstyle a'\xi (q^2-M^2-\Pi_T^{WW})}{\textstyle
q^2(1-\xi)+
               \xi (\Pi_T^{WW}-\Pi_L^{WW})}
\vspace{.5cm}
\end{equation}
               where we have defined $a'=a/[(1-a)(q^2-\xi M^2-\xi
\Pi_L^{WW})]$. Thus, the
               would-be goldstones will not give rise to a pole in the
$W$-boson
               propagator (see below).

               If we neglect the terms proportional to $a'$ which are
$q_{\mu}q_{\nu}$ terms
               of ${\cal O}(g^4(q^2-M^2-\Pi_T^{WW})^0)$ ---where $g$ is the
               SU(2) gauge coupling constant---, Eq. (4) becomes:
\vspace{.5cm}
  \begin{equation}
               P_{\mu\nu}(q)=\frac{\textstyle i}{\textstyle q^2-M^2-\Pi_T^{WW}}
               \left \{ -g_{\mu\nu}+\frac{\textstyle q_{\mu}q_{\nu}}{\textstyle
q^2} \cdot
               \frac{\textstyle q^2(1-\xi ) +\xi
(\Pi_T^{WW}-\Pi_L^{WW})}{\textstyle q^2 -
               \xi M^2 -\xi \Pi_L^{WW}}\right \} .
\vspace{.5cm}
  \end{equation}

               Observe that if we use Eq.(3b) into Eq.(6) we obtain the result
given in
             Eq.(20) of ref.[7], namely:
\vspace{.5cm}
\begin{eqnarray*}
             P_{\mu\nu}^{[7]}(q)=\frac{\textstyle i}{q^2-M^2+\Pi_T} \left \{
-g_{\mu\nu}
             +\frac{\textstyle q_{\mu}q_{\nu}(1-\xi-\xi \Pi_L)}{\textstyle q^2
-
             \xi (M^2-\Pi_T+q^2\Pi_L)}\right \}.
\vspace{.5cm}
\end{eqnarray*}

           Instead, if we choose to work in the unitary gauge, Eqs.(6) and (3c)
give rise to
             Eq.(9) of ref. [8]:
\vspace{.5cm}
\begin{eqnarray*}
             P_{\mu\nu}^{[8]}(q)= i\frac{ \left [\textstyle
-g_{\mu\nu}+q_{\mu}q_{\nu}
             \left ( \frac{\textstyle 1-F_2}{\textstyle M^2+F_1} \right )
\right ] }
             {\textstyle q^2-M^2-F_1-q^2 F_2},
             \end{eqnarray*}
             while using Eqs.(6) and (3d) one obtains in the unitary gauge:
             \begin{eqnarray*}
             P_{\mu\nu}^{[9]}(q)=i\frac{ \left [\textstyle
-g_{\mu\nu}+\frac{\textstyle
             q_{\mu}q_{\nu}}{\textstyle q^2} \left (\frac{ \textstyle q^2 +
             \overline{\epsilon}_T +
             \overline{\epsilon}_L}{\textstyle M^2+\overline{\epsilon}_L}\right
)
             \right ]}{\textstyle q^2-M^2+\overline{\epsilon}_T}.
\vspace{.5cm}
\end{eqnarray*}
             which reproduces Eq.(3) of ref.[9] when one takes the imaginary
parts
             of $\overline{\epsilon}_{T,L}$.

             The derivation of the last three equations starting from (6) shows
that
             , before renormalization, the results of refs. [7,8,9] are
equivalent since
             they differ only by the parametrizations used for the
self-energies, Eqs.(3).

             Now, Eq.(6) can be rewritten into the following form:
\vspace{.5cm}
\begin{equation}
             P_{\mu\nu}(q)=i\left \{ \frac{\textstyle -g_{\mu\nu}+
             \frac{\textstyle q_{\mu}q_{\nu}}{\textstyle q^2}}
             {\textstyle q^2-M^2-\Pi_T^{WW}}-\frac{\textstyle q_{\mu}q_{\nu}}{
             \textstyle q^2} \cdot \frac{\textstyle \xi }{\textstyle q^2 -\xi
M^2 -
             \xi \Pi_L^{WW}} \right \}
\vspace{.5cm}
\end{equation}
             which means that only the transverse part of the propagator will
             develop a $S$-matrix pole after renormalization.

               Following references [1--5] we replace the bare mass in Eq.(6)
in
               terms of the pole in the propagator:
\vspace{.5cm}
  \begin{eqnarray}
               \widetilde{M}^2 &\equiv & M_W^2 - iM_W \Gamma_W  \\
                   & = & M^2 + \overline{\Pi}_T^{WW}.
\vspace{.5cm}
   \end{eqnarray}
             (In the following $\overline{\Pi}^{WW}_i$ is used for $\Pi_i^{WW}
             (\widetilde{M}^2)$.)

    Next, we can expand around the pole $\widetilde{M}^2$ the denominator
    in Eq.(6); we obtain:
\vspace{.5cm}
\begin{eqnarray}
    q^2-M^2-\Pi_T^{WW}(q^2) &=& q^2-\widetilde{M}^2 -\Pi_T^{WW}(q^2)+
    \overline{\Pi}_T^{WW} \nonumber \\
    & =& Z^{-1}(q^2-\widetilde{M}^2)\{1+{\cal O}(g^2(q^2-\widetilde{M}^2)) \}.
\vspace{.5cm}
\end{eqnarray}
    where $Z^{-1} \equiv 1-\overline{\Pi'}^{WW}_T$.

     If we also expand the coefficient of $q_{\mu}q_{\nu}$ ---the term within
     curly brackets in Eq.(6)---, we get:
\vspace{.5cm}
\begin{eqnarray}
    \frac{1}{\textstyle q^2} \cdot \frac{\textstyle q^2(1-\xi)+\xi \Delta}
    {\textstyle q^2-\xi(M^2+\Pi^{WW}_L)} &=& \nonumber \\
    &=& \frac{1}{\textstyle \widetilde{M}^2} \left \{ 1 - \frac{\textstyle
    \widetilde{M}^2(1-\xi \overline{\Pi}^{WW}_T)+\xi \overline{\Delta}}
    {\textstyle \widetilde{M}^2(1-\xi) +
    \xi \overline{\Delta}} \left (\frac{\textstyle
    q^2 - \widetilde{M}^2}{\textstyle \widetilde{M}^2} \right ) +\cdots
\right\}
     \nonumber \\
     &\simeq & \frac{\textstyle 1-\xi'}{\textstyle q^2-\xi' \widetilde{M}^2}
\end{eqnarray}
    where $\Delta \equiv \Pi^{WW}_T-\Pi^{WW}_L$
    and
   \vspace{.5cm}
    \begin{eqnarray*}
    \xi' &\equiv& \frac{\textstyle (1-\overline{\Pi'}^{WW}_T) \xi}
    {\textstyle 1+\xi \left (\frac{\textstyle \overline{\Delta}}
    {\textstyle \widetilde{M}^2} -\overline{\Pi'}^{WW}_T \right )} \\
    \\ &=& Z^{-1} \xi [1+{\cal O}(g^2)]
\vspace{.5cm}
\end{eqnarray*}
    (see ref.[16]).

      Finally, if we replace Eqs.(10) and (11) into Eq.(6) we get:
\vspace{.5cm}
\begin{equation}
      P_{\mu\nu} = iZ \left \{ \frac{\textstyle -g_{\mu\nu} + (1-\xi')
      \frac{\textstyle q_{\mu}q_{\nu}}{\textstyle q^2-\xi' \widetilde{M}^2}}
      {\textstyle q^2 - \widetilde{M}^2} +\cdots \right \}
\vspace{.5cm}
\end{equation}
      where again, the ellipsis denote non-resonant $q_{\mu}q_{\nu}$ terms of
      ${\cal O}(g^2)$.

      Let us make some remarks about our results:

      (i) As it has been already discussed in refs. [1,4], the complex
constants
      $\Pi_T^{WW}(\widetilde{M}^2)$ in Eq.(9) and $Z$ in Eq.(12) play the role
      of the mass counterterm and $W$-field renormalization, respectively. In
the
      on-shell scheme [10], the corresponding quantities are real constants.
One
      can easily go from one scheme to the other by neglecting terms of
      ${\cal O}(g^4)$ (see Appendix A  and ref.[4]).

      Note that to reach
      the result in Eq.(12) one must renormalize the $\xi$ gauge
      parameter (see ref.[16]). Observe also that in order to correctly
      drawn the (lowest order) resonant propagator from the renormalized one,
      it is essential to extract the wavefunction renormalization constant $Z$.

      (ii)  The term inside the curly brackets in Eq.(12) contains the unique
       pole of the renormalized propagator. The remaining (non-resonant) terms
        are of ${\cal O}(g^2)$. Thus, {\em at lowest order} we recover the
        $W^{\pm}$ propagator in the resonance region for an arbitrary $\xi$-
        gauge. In other words, the lowest order propagator at the resonance is
        obtained by replacing $M^2$ in Eq.(2) by the complex pole
$\widetilde{M}^2$
         as already stated in our previous paper [3].

         (iii) If we formally expand $\Pi_T^{WW}(\widetilde{M}^2)$ around the
mass
         $M_W^2$ ---Eq.(8)--- and then compare the real and imaginary parts of
         Eqs.(8,9), we
         obtain the following relations:
\vspace{.5cm}
 \begin{eqnarray}
         M_W^2 &=& M^2 + Re\Pi_T^{WW}(M_W^2) + \cdots \\
         -M_W\Gamma_W &=& Im\Pi_T^{WW}(M_W^2) + \cdots
\vspace{.5cm}
\end{eqnarray}
         where the ellipsis denote terms of ${\cal O}(g^4)$ since $\Gamma_W$
         starts at ${\cal O}(g^2)$. The above expansion reproduces the on-shell
         renormalized mass, Eq.(13), and the unitarity relation (14) when taken
         at leading order; note however that unitarity is a relation exactly
         valid order by order. It should be noted also that $M_W^2$ above
denotes
         the real part of the pole position and not the renormalized mass in
the
         on-shell scheme.

        (iv) Finally, the unitarity and mass renormalization relations of
         ref.[8] are obtained by using Eq.(3c) and the leading order
expressions
         for Eqs.(13,14).
         On the other hand, since in ref.[9] only the imaginary parts of the
$W$ self-
         energy has been included in the Dyson summation, only Eq.(14) is
         reproduced as it can be explicitly shown from the expresion
         $\overline{\epsilon}_T
         \sim\! i\Gamma^0_W$ given in ref.[9].

          Summarizing, in this paper we have shown that the lowest order
          $W^{\pm}$ propagator in the resonant region is given in an
          arbitrary $\xi$
          gauge by the term inside the curly brackets in Eq.(12), i.e.:
\vspace{.5cm}
\begin{equation}
          \overline{\Delta}_{\mu\nu}(q) = \frac{i \left \{ \textstyle
-g_{\mu\nu}
          +(1-\xi ) \frac{\textstyle q_{\mu}q_{\nu}}{\textstyle q^2 -
          \xi (M_W^2-iM_W\Gamma_W)} \right \} }
          {q^2 - M_W^2 + iM_W \Gamma_W}.
\vspace{.5cm}
\end{equation}

          We have shown that this propagator is the leading term in the
expansion of the
          renormalized propagator around the pole $\widetilde{M}^2$ and that he
non-resonant
          terms are explicitly of ${\cal O}(g^2)$.
          The unique pole of the propagator in Eq.(15) is located in its
          transverse part as it can be shown by projecting out this equation.
          Finally, the Ward identity used in ref.[3] is consistent since it
          involves only lowest order quantities.

\

          {\bf Acknowledgements}

    One of us (GLC) acknowledges partial finantial support from CONACyT.
    The work of JLLM has been partially supported by CONACyT under contracts
    F246-E9207 and 1628-E9209.

        \

\newpage

\begin{center}
          {\bf Appendix A}
\end{center}
           In this appendix we discuss the relationship between the
wavefunction
           renormalization in the on-shell scheme [10] and the corresponding
           quantity when we fix the pole in the propagator in the $S$-matrix
pole
           scheme [1,4].
            As we argued above in the text, one can move from one to the other
scheme by neglecting
            terms of ${\cal O}(g^4)$.

              For definiteness we consider the complete propagator for an
scalar particle.
\vspace{.5cm}
\begin{eqnarray*}
               \frac{1}{D} = \frac{1}{\textstyle q^2 - m_0^2 - \Pi (q^2)}
              \hspace{5.5cm} (A1)
\vspace{.5cm}
  \end{eqnarray*}
               where $m_0$ denotes its bare mass and $\Pi (q^2)$ its 1PI
self-energy.

                The on-shell renormalization [10] is obtained by expanding
$Re\Pi (q^2)$
                around the renormalized mass $m_R^2$:
\vspace{.5cm}
   \begin{eqnarray*}
                 D &=& q^2-m_0^2 - Re\Pi (m_R^2) - (q^2 - m_R^2)Re\Pi' (m_R^2)
+ \cdots
                 -i Im\Pi (q^2)  \hspace{0.5cm} (A2) \\
                   & \simeq & Z^{-1} \{ q^2 - m_R^2 - iZ Im \Pi (q^2) \}
                   \hspace{6.5cm} (A3)
\vspace{.5cm}
      \end{eqnarray*}
                   where $Z^{-1} \equiv 1- Re \Pi' (m_R^2)$ and $m_R^2 = m_0^2
+
                   Re\Pi (m_R^2)$ correspond to the on-shell wavefunction and
mass
                   renormalization
                   . Notice that both renormalized quantities, $Z^{-1}$ and
$m_R$,
                   are real.

                     If we choose to expand also $Im\Pi (q^2)$ in Eq.(A2)
around $m_R$ we
                     get:
\vspace{.5cm}
        \begin{eqnarray*}
                     {D} &=& \overline{Z}^{-1} \left \{ q^2 - m_R^2 - i
                     \overline{Z}Im\Pi
                     (m_R^2) \right \} \nonumber \\
                     &=& \overline{Z}^{-1} \left \{ q^2 - m_R^2 + im_R \Gamma_R
\right \}
                     \hspace{3cm} (A4)
\vspace{.5cm}
        \end{eqnarray*}
                     where the second relation above follows from unitarity.

                     Observe that now the wavefunction renormalization
$\overline{Z}^{-1}=
                     1-\Pi'(m_R^2)$, becomes a complex quantity. This quantity
can be
                     rewritten as follows:
\vspace{.5cm}
        \begin{eqnarray*}
                     \overline{Z}^{-1} = Z^{-1} \left \{ 1- iIm \Pi'(m_R^2)
\right \} +
                     {\cal O}(g^4), \hspace{2cm} (A5)
\vspace{.5cm}
        \end{eqnarray*}
                     where $g$ is the relevant coupling constant of the scalar
particle
                     to the particles involved in the 1PI graph. Thus, up to
terms of
                     ${\cal O}(g^4)$, Eq.(A5) furnish the relationship between
                     Eqs.(A3) and (A4) (see also ref.[4]).

                        In contrast to the real renormalization constants in
the on-shell
                        scheme [10], the requirement of a defined $S$-matrix
pole in the propagator as
                        in Eq.(A4)
                        naturally involves a complex-valued wavefunction
renormalization.

 \

\begin{center}
\bf Appendix B
\end{center}

In this appendix we use a simple model to illustrate how a resonant
propagator with an energy-dependent width (for example the one of Ref.
[9]) leads to violations of gauge invariance.

  Let us consider the $s$-wave $\pi^+ \eta$ scattering: $\pi^+(p) \eta(q)
\rightarrow \pi^+(p')\eta(q')$. Near $\sqrt{s} \sim $ 1 GeV, this process
is dominated by the $a_0^-(980)$ meson. If we assume an energy-dependent
width for the $a_0$ propagator, the corresponding scattering amplitude
can be written as follows:
\vspace{.5cm}
\[
{\cal M}^0 = \frac{-ig^2}{s-m^2+im\Gamma(s)} \hspace{3cm} (B1)
\vspace{.5cm}
\]
where $s=(p+q)^2=(p'+q')^2$ is the squared of the center of mass energy and
$g$ the $a_0 \pi^+ \eta$ coupling constant.

  Now let us consider the corresponding radiative process: $\pi^+(p)
\eta(q) \rightarrow \pi^+(p') \eta(q') \gamma(\epsilon, k)$, where
$\epsilon$ and $k$ denote the four-polarization vector and four-momentum
of the photon. For simplicity we introduce the kinematical variables
$s=(p+q)^2,\ s'=(p'+q')^2$, such that $s=s' + 2(p+q)\cdot k$.

  The scattering amplitude receives contributions from three sources: the
emission of the $\gamma$ from the $\pi^+$ external lines and the emission
from the $a_0^-$ line. The explicit form of the amplitude is:
\vspace{.5cm}
\begin{eqnarray*}
{\cal M} = ieg^2 \epsilon^{\mu}&&\hspace{-1.0cm} \left\{
\frac{p_{\mu}}{p. k} \cdot
\frac{1}{s'-m^2+im\Gamma(s')} - \frac{p'_{\mu}}{p'. k} \cdot \frac{1}{s -
m^2 + im\Gamma(s)} \right. \\
&-& \left.  \frac{2(p+q)_{\mu}}{[s-m^2+im\Gamma(s)][s'-m^2+im\Gamma(s')]}
 \right\} \hspace{2cm} (B2)
\vspace{.5cm}
\end{eqnarray*}
where $e$ denotes the $\pi^+$ electric charge.

  As is well known, electromagnetic gauge-invariance requires that ${\cal
M}=0$ when $\epsilon \rightarrow k$. Thus, Eq. (B2) satisfies
gauge-invariance only if $\Gamma(s)=\Gamma(s')$, {\em i.e.} the width in
the propagator has to be a constant.

   In the same way, it is very easy to check that the amplitude for the
process $t \rightarrow b \tau^+ \nu_{\tau} \gamma$ is gauge invariant
under electromagnetism only if Eq. (1) is used for the propagator of the
virtual $W^+$.

\end{document}